\documentclass[a4paper,11pt]{article}
\pdfoutput=1

\usepackage[english]{babel}

\usepackage{jcappub}

\usepackage[T1]{fontenc}

\usepackage{mathtools}

\usepackage{amssymb}

\usepackage{url}
\usepackage{amsmath}
\usepackage{graphicx}
\graphicspath{{images/}}

\title{Stochastic isocurvature constraints for axion dark matter with high-scale inflation}

\author[]{Liina Jukko}
\author[]{and Arttu Rajantie}

\affiliation[]{Department of Physics, Imperial College London, London, SW7 2AZ, United Kingdom}

\emailAdd{liina.jukko@gmail.com}
\emailAdd{a.rajantie@imperial.ac.uk}

\abstract{ Axions are among the best motivated dark matter candidates. Their production in the early Universe by the vacuum misalignment mechanism gives rise to isocurvature perturbations, which are constrained by cosmic microwave background measurements. In this paper, we compute the axion isocurvature power spectrum using spectral expansion in the stochastic Starobinsky-Yokoyama formalism, which captures non-linear effects in the axion dynamics. In contrast to most of the existing literature, we focus on high inflationary Hubble rates of order $10^{13}~{\rm GeV}$, and demonstrate that there is a significant window in which axions can account for all or part of the dark matter abundance without violating the isocurvature bounds or tensor mode bounds. Crucially, we find that the isocurvature spectrum is dominated by non-perturbative contributions in a large part of this window. Therefore the commonly used linear approximation is not reliable in this region, making the stochastic approach essential.
}

\begin{document}
\maketitle
\flushbottom

\section{Introduction}  %

Axions~\cite{Wilczek:1977pj,Weinberg:1977ma} are hypothetical particles which arise as pseudo-Goldstone bosons when the Peccei-Quinn symmetry~\cite{Peccei:1977hh,Peccei:1977ur} is spontaneously broken as a 
solution to the strong CP problem.
As stable, neutral particles they are a natural candidate
for cold dark matter (CDM), and
a lot of theoretical work has been done in the past few decades on both conventional QCD axions and axion-like particles (ALPs)~\cite{Sikivie:2006ni,Wantz:2009it,Marsh:2015xka}. In this paper we assume that the Peccei-Quinn symmetry undergoes phase transition during inflation, producing axions through the vacuum misalignment mechanism (see e.g. Ref. \cite{Preskill:1982cy,Abbott:1982af,Dine:1982ah,Turner:1985si}), where the coherent oscillations of the axion field in the post-inflationary epoch account for the axion CDM particle production. 

\par Quantum fluctuations in the axion field during the inflationary epoch introduce an isocurvature component to the Cosmic Microwave Background (CMB)~\cite{Steinhardt:1983ia,Seckel:1985tj,Lyth:1989pb,Hertzberg:2008wr,Hamann:2009yf, Wantz:2009it}, which is heavily constrained by current observations \cite{Akrami:2018odb}. 
Most of the existing work has focussed on the ``anthropic window''~\cite{Hertzberg:2008wr} at low inflationary Hubble rates, $H\lesssim 10^8~{\rm GeV}$, where the axion field is almost uniform over the whole currently observable Universe.
Instead, we will focus on the ``classic window'' identified in Ref.~\cite{Hertzberg:2008wr} at $H\sim 10^{13}~{\rm GeV}$, which has received less attention in the context of isocurvature perturbations.
As we will see, the usual linear or perturbative techniques are not fully applicable in this case, and instead, we will use the stochastic approach~\cite{Starobinsky:1986fx,Starobinsky:1994bd}.

In the stochastic approach, the dynamics of the axion field on super-horizon scales is described by a stochastic Langevin equation. It has been applied to axions before, e.g., in Refs.~\cite{Graham:2018jyp,Nakagawa:2020eeg}, where it was used to compute the one-point probability distribution of the field. In this paper, we go beyond that by using the spectral expansion method~\cite{Starobinsky:1994bd,Markkanen:2018gcw,Markkanen:2019kpv,Markkanen:2020bfc}, which allows the numerical determination of the exact asymptotic long-distance behaviour of any correlation function in de Sitter spacetime within the stochastic approximation. This allows us to compute the isocurvature power spectrum without having to rely on linear or perturbative assumption.

This paper is organised as follows. In Section~\ref{approach}, we introduce the axion dark matter model and the stochastic approach. In Section~\ref{results}, we present the results of our calculations and the observational constraints that follow from them. We then discuss our findings and present our conclusions in Section~\ref{discussion}.

\section{Stochastic approach for the axion} \label{approach}

\subsection{Axion dark matter} \label{PQsoln}

Let us consider an axion field $a$ with cosine potential
\begin{equation} \label{eq: actualaxionpot}
V(a) = V_0 \bigg[ 1 - \cos{\bigg( \frac{a}{F_a}  \bigg)} \bigg],
\end{equation}
where $F_a$ is the decay constant of the axion, and $V_0$ is a constant which is related to the axion mass $m_a$ through $V_0=F_a^2m_a^2$.
The axion field $a(x)$ can be written as
 \begin{equation} \label{eq: axionfieldwithphi}
 a(x) = F_a \phi(x),   
\end{equation}
 where $\phi(x)$ is a dimensionless phase angle.

We will consider both a general axion scenario, in which the particles are often referred to as ALPs and in which $V_0$ and $F_a$ are free parameters, and the specific case of the QCD axion, in which the potential is generated by non-perturbative QCD effects, so that
\begin{equation} \label{eq: v0ofqcdaxion}
V_0 \simeq \Lambda_{QCD}^4,    
\end{equation}
where $\Lambda_{QCD}\sim 250~{\rm MeV}$ is the QCD breaking scale. %
Although it is common to use the term axion to refer to only QCD axions, we use is cover both cases for simplicity.

As stable, electrically neutral scalar particles, axions are a candidate for dark matter. We will focus on the scenario in which they are produced by the vacuum misalignment mechanism~\cite{Preskill:1982cy,Abbott:1982af,Dine:1982ah,Turner:1985si}: During inflation, quantum fluctuations drive the axion field away from its vacuum state so that when inflation ends, the phase angle $\phi$ has a non-zero value, known as the vacuum misalignment angle.
Eventually the axion field starts to oscillate about its vacuum state, and these oscillations can be interpreted as dark matter particles. 
The energy density $\rho$ of these particles depends on the amplitude of the oscillations, which is determined by the vacuum misalignment angle. 

If the potential was purely harmonic, the axion field would simply undergo damped harmonic oscillations, and the produced dark matter density would be simply proportional to the potential energy,
\begin{equation}
\label{equ:harmonic}
 \rho_{\rm harm}(\phi)  \propto \frac{1}{2}V_0\phi^2.
\end{equation}
However, for the cosine potential (\ref{eq: actualaxionpot}) the dependence is more complicated~\cite{Turner:1986tb}, which is easy to see by considering a misalignment angle $\phi\approx\pi$, which is close to the top of the potential barrier. In that case the field will initially roll very slowly away from the maximum, and during this time the energy density remains approximately constant. The final energy density is therefore much higher than what the harmonic approximation would predict. Including this effect, the final energy density can be approximated by~\cite{Lyth:1991ub,Marsh:2015xka} 
\begin{equation} \label{eq: thefunction}
\rho(\phi) \propto V_0 f(\phi),\quad f(\phi)= \frac{\phi^2}{2} \bigg[\ln{\frac{e}{1-\frac{\phi^2}{\pi^2}}} \bigg]^\frac{7}{6},
\end{equation}
where we have introduced the dimensionless anharmonic function $f(\phi)$.

The correlation length of the misalignment angle $\phi$ depends greatly on the parameters. We will discuss this in more detail in Section~\ref{results}, but as a rule of thumb, if $H\ll F_a$ then the correlation length is many Hubble lengths. In that case, the misalignment angle $\phi$ can be considered to be uniform across the whole Universe, with a random value that cannot be predicted. The dark matter density parameter $\Omega_{DM}$ would be determined by it through Eq.~(\ref{eq: thefunction}), and one can ask how likely it is to agree with the observed dark matter density parameter
\cite{Planck:2018vyg},
\begin{equation} \label{eq: dmbound}
\Omega_{DM}h^2 \simeq 0.12 \: .
\end{equation}
This leads to anthropic considerations, and therefore this is often called the "anthropic window"~\cite{Hertzberg:2008wr,Wantz:2009it}.
This is the case most existing literature on isocurvature perturbations has focused on.

In contrast, we consider the opposite case $H\gtrsim F_a$, in which the currently observable Universe consists of a large number of patches between which the misalignment angle $\phi$ was uncorrelated. Locally in each post-inflationary Hubble volume, Eq.~(\ref{eq: thefunction}) is still valid, but observations should be compared with its average value
\begin{equation}
\label{equ:Omegadm}
\Omega_{DM}h^2 = 0.12 \times 2 \langle f(\phi)\rangle  \bigg( \frac{F_a}{5 \times 10^{11} \: \textrm{GeV}} \bigg)^{\frac{7}{6}},
\end{equation}
where the expectation value $\langle f(\phi)\rangle$ is the average over the equilibrium distribution of the field at the end of inflation.
Because Eq.~(\ref{eq: thefunction}) is not a linear function, this is not the same as the dark matter density parameter computed using the average misalignment angle. 

The position-dependence of the misalignment angle also means that the produced dark matter density will be different in different parts of the Universe, which corresponds to isocurvature perturbations and can be potentially observed through measurements of the cosmic microwave background (CMB) anisotropy.
The local isocurvature perturbation $S(\vec{x})$ is given by the relative fluctuation of the dark matter density $\rho$,
\begin{equation}
    S(\vec{x})=\frac{\rho(\vec{x})-\langle \rho\rangle}{\langle\rho\rangle}.  
\end{equation}
CMB observations place an upper bound on the power spectrum ${\cal P}_S$ of this quantity, 
given by the Fourier transform of the correlation function
\begin{equation}
\label{equ:isocorr}
       G_S(\vec{x})\equiv\langle S(0)S(\vec{x})\rangle
    =\frac{\langle \rho(0)\rho(\vec{x})\rangle}{\langle\rho\rangle^2}-1,
\end{equation}
as
\begin{equation}
{\cal P}_S(k)\equiv \frac{k^3}{2\pi^2}
\int d^3x\, e^{i\vec{k}\cdot\vec{x}}G_S(\vec{x}).
\end{equation}
The current observational constraint is~\cite{Akrami:2018odb}
\begin{equation} \label{eq: bound}
{\cal P}_S(k_P) \lesssim 0.040 {\cal P}_{\zeta}(k_P)\approx  8.8 \times 10^{-11}
\end{equation}
at the Planck pivot scale 
\begin{equation}
\label{equ:Planckpivot}
    k_P=0.05~{\rm Mpc}^{-1}.
\end{equation}
This places constraints on axion dark matter models. The isocurvature power spectrum predicted by axion dark matter scenarios has been studied in the literature, but usually in the harmonic approximation. In what follows, we compute it %
non-perturbatively using the stochastic approach.

\subsection{The stochastic method}

In de Sitter spacetime, which is a good approximation for spacetime geometry during inflation, quantum fluctuations of any light scalar field are amplified by the expansion of space. On superhorizon scales, it can therefore be considered as a classical field. However, as comoving quantum modes leave the horizon, they contribute to this classical field, giving rise to a stochastic noise term. Therefore the superhorizon dynamics of the scalar field is described by a stochastic classical Langevin equation~\cite{Starobinsky:1986fx,Starobinsky:1994bd},
\begin{equation}
    \dot{\phi}=-\frac{1}{3H F_a^2}\frac{\partial V}{\partial \phi}+\xi,
\end{equation}
and $\xi$ is a Gaussian stochastic white noise term, which satisfies
\begin{equation}
    \langle\xi(t)\xi(t')\rangle=\frac{H^3}{4\pi F_a^2}\delta(t-t').
\end{equation}
Here $H$ is the Hubble rate during inflation, which we assume to be constant. The absence of primordial tensor modes in CMB observations implies the upper bound~\cite{Akrami:2018odb}
\begin{equation}
    \label{equ:Hbound}
    H\lesssim 6.1\times 10^{13}~{\rm GeV}.
\end{equation}

The time evolution of the probability distribution $P(\phi;t)$ then satisfies the Fokker-Planck equation
\begin{equation}
    \frac{\partial P(\phi;t)}{\partial{t}}=
    \frac{1}{3HF_a^2}\frac{\partial^2 V}{\partial\phi^2}P+
    \frac{1}{3HF_a^2}\frac{\partial V}{\partial\phi}\frac{\partial P}{\partial \phi}
    +\frac{H^3}{8\pi^2 F_a^2}\frac{\partial^2 P}{\partial \phi^2}.
\end{equation}
This can be solved using the spectral expansion as
\begin{equation}
P(\phi;t)=e^{-\frac{4\pi^2}{3H^4}V(\phi)}\sum_n c_n\psi_n(\phi)
e^{-\Lambda_n t},
\end{equation}
in terms of arbitrary coefficients $c_n$ and eigenfunctions $\psi_n(\phi)$ and eigenvalues $\Lambda_n$,
which satisfy the Schrödinger-like equation~\cite{Starobinsky:1994bd,Markkanen:2019kpv}
\begin{equation} \label{eq: eigeneq}
D_{\phi}\psi_n(\phi) = -\frac{4\pi^2F_a^2\Lambda_n}{H^3}\psi_n(\phi),
\end{equation}
with
\begin{equation} \label{eq: tildestuff}
D_{\phi} = \frac{1}{2}\frac{\partial^2}{\partial\phi^2}
- \frac{1}{2}\big(v'(\phi)^2 - v''(\phi)\big) \;\; \textrm{and} \;\; v(\phi) = \frac{4\pi^2}{3H^4} V(\phi).
\end{equation}
The eigenfunctions $\psi_n(\phi)$ are assumed to be real and complete, and orthonormal such that
\begin{equation}
    \int d\phi\, \psi_n(\phi)\psi_{n'}(\phi)=\delta_{n,n'}.
\end{equation}
One can also show (see \cite{Starobinsky:1994bd}) that $\Lambda_n \geq 0$. It can easily be seen that a stationary solution exists, 
\begin{equation}
    \psi_0(\phi)\propto \exp\left(-\frac{4\pi^2 V(\phi)}{3H^4}\right),
\end{equation}
i.e. the lowest eigenvalue is $\Lambda_0 = 0$.
This gives the well-known equilibrium one-point probability distribution~\cite{Starobinsky:1994bd}
\begin{equation}
\label{equ:eqPDF}
    P_{\rm eq}(\phi)=\psi_0(\phi)^2
    \propto \exp\left(-\frac{8\pi^2 V(\phi)}{3H^4}\right).
\end{equation}
Using this, it is possible to calculate the expectation value $\langle f(\phi)\rangle$ in Eq.~(\ref{equ:Omegadm})
as
\begin{equation}
\label{equ:expvalue}
    \langle f(\phi)\rangle=\int d\phi\,P_{\rm eq}(\phi)f(\phi).
\end{equation}

The spectral expansion also makes it possible to compute correlation functions, such as Eq.~(\ref{equ:isocorr}).
The correlator of any function $f(\phi)$ between two different times $t_1$ and $t_2$ is given by~\cite{Starobinsky:1994bd,Markkanen:2019kpv}
\begin{equation}
    \left\langle f\big(\phi(t_1)\big) f\big(\phi(t_2)\big)\right\rangle=
    \sum_n f_n^2 e^{-\Lambda_n |t_2-t_1|},
\end{equation}
where the spectral coefficients $f_n$ are
\begin{equation} \label{eq: coeff-f}
f_n = \langle 0 | f | n \rangle = \int d\phi\, \psi_0(\phi)f(\phi)\psi_n(\phi).
\end{equation}
For the isocurvature power spectrum (\ref{equ:isocorr}) we need the spatial equal-time correlator, which can be obtained using de Sitter invariance~\cite{Starobinsky:1994bd,Markkanen:2019kpv} as
\begin{equation}\label{eq: spatialcorrelator}
     \left\langle f\big(\phi(0)\big) f\big(\phi(\vec{x})\big)\right\rangle=\sum_{n=0}^{\infty} f_n^2 \frac{1}{(|\vec{x}|H)^{\frac{2\Lambda_n}{H}}}.
\end{equation}

Specifically, the two-point correlator of the isocurvature perturbation (\ref{equ:isocorr}) is therefore given by
\begin{equation} \label{eq: finalisocurvcorr}
\langle S(0) S(\vec{x}) \rangle = \sum_{n=1}^{\infty} \frac{f_n^2}{\langle f \rangle^2} \frac{1}{(|\vec{x}|H)^{\frac{2\Lambda_n}{H}}}, 
\end{equation}
and its Fourier transform gives the power spectrum
\begin{equation}
\label{equ:PS}
{\cal P}_S(k) \approx \frac{2}{\pi}  \sum_{n=1}^\infty \frac{f_n^2}{\langle f \rangle^2} \Gamma\bigg( 2 - \frac{2\Lambda_n}{H} \bigg) \sin{\bigg( \frac{\Lambda_n \pi}{H} \bigg)} \bigg( \frac{a_0k}{a_{\rm inf}H} \bigg)^\frac{2 \Lambda_n}{H},
\end{equation}
where $a_0$ and $a_{\rm inf}$ are the scale factors today and at the end of inflation, respectively. 
For the Planck pivot scale (\ref{eq: bound}), the quantity in the last brackets is very small,
\begin{equation}
\frac{a_0k_P}{a_{\rm inf}H}\equiv
e^{-N_P}\approx \bigg( \frac{H}{8 \times 10^{13} \: \textrm{GeV}} \bigg)^{-1/2} e^{-56}.
\end{equation}
Therefore this sum is dominated by its first few terms, and we can compute the power spectrum reliably by finding the first few eigenfunctions $\psi_n(\phi)$ and the corresponding eigenvalues $\Lambda_n$.

\subsection{Axion spectrum} \label{findingevalueequationforaxion}

\begin{figure} %
\centering
\includegraphics[scale=0.9]{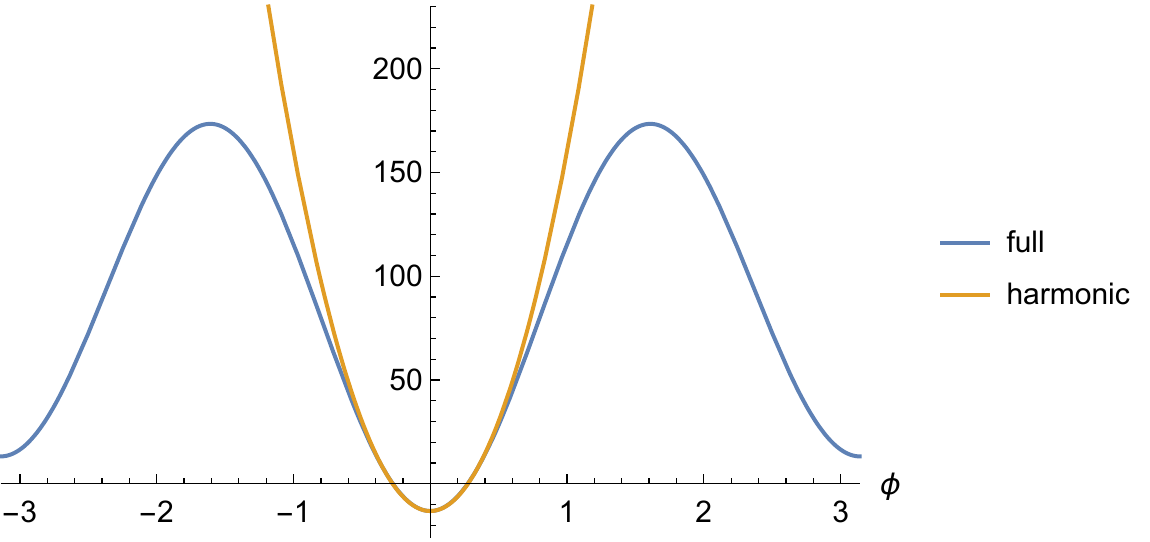}
\caption{The ``potential'' $W(\phi)$ entering the Schrödinger type eigenvalue equation \eqref{eq: simpleeveq} for $\alpha = 1$. The blue curve shows the full function, while the orange curve shows the harmonic approximation.}
\label{fig: SEpot}
\end{figure}

For the axion potential (\ref{eq: actualaxionpot}),
the eigenvalue equation (\ref{eq: eigeneq}) has the form
\begin{equation} \label{eq: simpleeveq}
\frac{\partial^2 \psi_n}{\partial \phi^2} 
-  \bigg[ \frac{16\pi^4}{9} \alpha^2 \sin^2{\phi}  - \frac{4\pi^2}{3} \alpha \cos{\phi} \bigg] \psi_n 
= -8\pi^2 \beta_n \: \psi_n,  
\end{equation}
where we have introduced the dimensionless parameters
\begin{equation} \label{eq: beta}
\alpha = \frac{V_0}{H^4},\quad
\beta_n = \frac{F_a^2 \Lambda_n}{H^3}.
\end{equation}
It is important to note that this equation, and therefore the spectrum, depends only on $\alpha$.

Equation (\ref{eq: simpleeveq}) has the form of the Schr\"odinger equation with ``potential'' $W(\phi)=v'(\phi)^2-v''(\phi)$,
shown in Fig.~\ref{fig: SEpot} for $\alpha=1$.
It has a minimum at $\phi=0$, which is also the minimum of the axion potential $V(\phi)$, and if $\alpha>3/8\pi^2$, it has also another higher minimum at $\phi=\pm\pi$, which is a maximum of the axion potential. 

Because $\phi$ is assumed to be a phase angle, the eigenfunctions $\psi_n$ have to be periodic. Therefore Eq.~(\ref{eq: eigeneq}) is defined over the range $-\pi<\phi\le \pi$, with periodic boundary conditions. The equation has reflection symmetry $\phi\rightarrow -\phi$, which means that the eigenfunctions can be divided into even and odd parity.
Furthermore, in general, the eigenvalue equation (\ref{eq: eigeneq}) has the form of the Schrödinger equation in supersymmetric quantum mechanics~\cite{Cooper:1982dm,Cooper:1994eh}. The supersymmetry transformation 
\begin{equation}\label{equ:susy}
    \psi \rightarrow \tilde\psi=\psi'+v'\psi
\end{equation}
maps eigenfunctions of the original potential $V(\phi)$ to those of the flipped potential $-V(\phi)$ with the same eigenvalue. In the case of Eq.~(\ref{eq: actualaxionpot}), this sign flip can be cancelled by shifting the field $\phi\rightarrow\phi+\pi$, and therefore the supersymmetry transformation is actually a map to a different eigenfunction in the same theory.  Because the supersymmetry transformation changes the parity, the spectrum consists of degenerate pairs of odd and even eigenfunctions, i.e.,
$\Lambda_{2n}=\Lambda_{2n-1}$ for all $n>0$. With correction terms for the axion potential, this shift symmetry is broken and the eigenspectrum loses its degeneracy. 

\subsection{Harmonic approximation} \label{harmandmixedapproxs}

For comparison with the full stochastic treatment for the axion, we first consider the harmonic approximation, in which the  axion potential \eqref{eq: actualaxionpot} has the quadratic form. We use

\begin{equation} \label{eq: harmapproxpot}
V(\phi) = \frac{1}{2}V_0\phi^2    
\end{equation}

\noindent for the potential, and the quadratic energy density (\ref{equ:harmonic}), which corresponds to
\begin{equation} \label{eq: quadratic}
f_{\textrm{harm}}(\phi) = \frac{\phi^2}{2}.   
\end{equation}

\noindent 
This corresponds to the free field approximation, and therefore the calculations could be done directly in quantum field theory without the use of the stochastic approximation. However, we use the stochastic approximation, because it works well as long as the axion mass is not much higher than the Hubble rate and because it allows a direct comparison with the full stochastic results.
Then following the steps in Section \ref{findingevalueequationforaxion} to find the eigenvalue equation for the quadratic potential \eqref{eq: harmapproxpot}, we get the harmonic eigenvalue equation, 
\begin{equation} \label{eq: harmapproxevalueeq}
\frac{\partial^2 \psi^{\rm harm}_n}{\partial \phi^2} 
-  \bigg[ \frac{16\pi^4}{9} \alpha^2 \phi^2  - \frac{4\pi^2}{3} \alpha \bigg] \psi^{\rm harm}_n 
= -8\pi^2 \beta^{\rm harm}_n \psi^{\rm harm}_n,  
\end{equation}
with $\alpha$ and $\beta^{\rm harm}_n$ defined by Eq.~\eqref{eq: beta} as before. 
This is the Schr\"odinger equation for the harmonic oscillator, and therefore the eigenfunctions and eigenvalues are well known,
\begin{equation}
\label{equ:harmef}
\psi^{\rm harm}_n(\phi)=\frac{1}{\sqrt{2^n n!}}\left(\frac{4\pi\alpha}{3}\right)^{1/4} H_n\left(\sqrt{\frac{4\pi^2 \alpha}{3}}\phi\right)
e^{-\frac{2\pi^2\alpha}{3}\phi^2},\quad 
\beta^{\rm harm}_n=\frac{n\alpha}{3},
\end{equation}
where $H_n$ are Hermite polynomials.
These give the expectation value and the spectral coefficients as
\begin{equation}
    \langle f_{\rm harm}\rangle =\frac{3}{16\pi^2\alpha},\quad
    f^{\rm harm}_2=\frac{3}{8\sqrt{2}\alpha\pi^2},
\end{equation}
with $f^{\rm harm}_n=0$ for $n=1$ and $n>2$.

\section{Results} \label{results}

\subsection{Parameters} \label{sec:constraints}

\begin{figure} %
\centering
\includegraphics[scale=0.9]{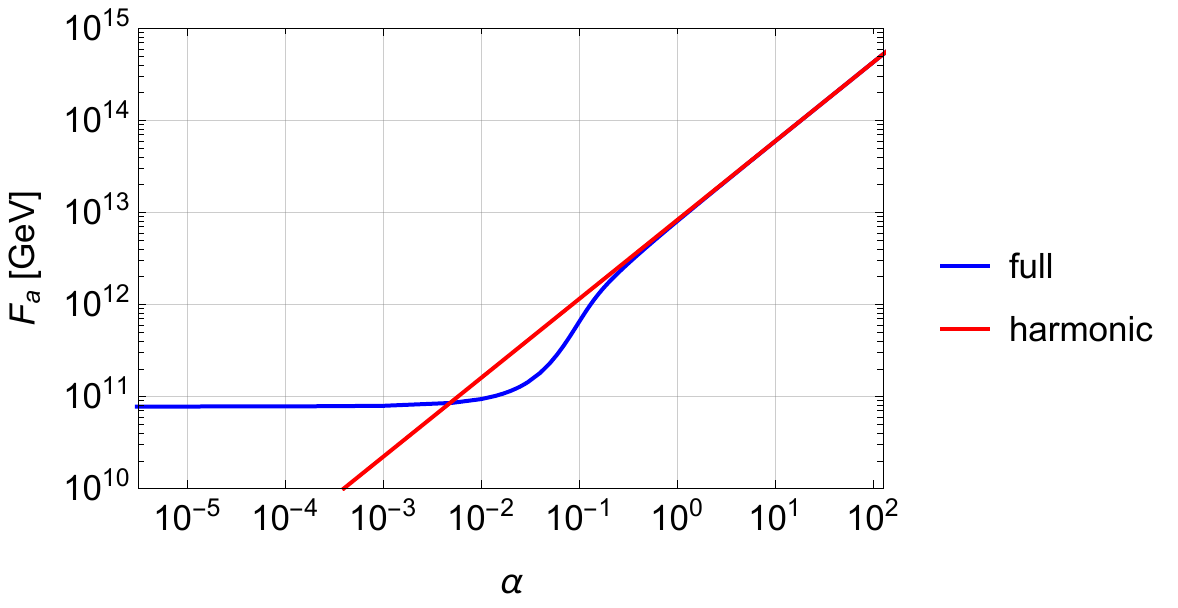}
\caption{The axion decay constant $F_a$ required to explain the observed dark matter abundance, as a function of the dimensionless $\alpha$ parameter \eqref{eq: beta}. 
The red line shows the harmonic approximation, and the blue curve shows the full stochastic result.
}
\label{fig: Faandalpha}
\end{figure}

The full parameter space of the general axion model is three-dimensional, consisting of the axion decay constant $F_a$, the inflationary Hubble rate $H$, and the dimensionless parameter $\alpha$ defined in Eq.~(\ref{eq: beta}). We focus on the case in which axions give a dominant contribution to the dark matter density, and therefore Eqs.~(\ref{eq: thefunction}), (\ref{eq: dmbound}) and (\ref{equ:Omegadm}) allow us to determine the required value of $F_a$,
\begin{equation}
\label{equ:dmrelation}
F_a\approx \frac{5\times 10^{11}~{\rm GeV}}{ \left(
2\left\langle f\right\rangle\right)^{6/7}}.
\end{equation}
The right-hand side, which can be computed using Eq.~(\ref{equ:expvalue}), depends monotonically on $\alpha$ as shown in Fig.~\ref{fig: Faandalpha}. Therefore we are left with a two-dimensional parameter space $(F_a,H)$. 

In the harmonic approximation, Eq.~(\ref{equ:dmrelation}) gives the relation $F_a\propto \alpha^{6/7}$. This would mean that for any value of $F_a$ it would be possible to find $\alpha$ that gives the correct dark matter density.
However, this is actually not true.
In the full case,  the highest possible value of $\langle f(\phi)\rangle$ is approximately $4.386$, reached at $\alpha=0$. This means that, as shown in Fig.~\ref{fig: Faandalpha}, axions can only be the dominant dark matter component if 
\begin{equation}
    \label{equ:abundance}
F_a\gtrsim 7.8\times 10^{10}~{\rm GeV}.
\end{equation}

In the specific case of the QCD axion, Eq.~(\ref{eq: v0ofqcdaxion}) gives another relation between the parameters. For the Hubble rates that we are interested in, this corresponds simply to the limit $\alpha\rightarrow 0$, and therefore one needs $F_a\approx 7.8\times 10^{10}~{\rm GeV}$ to get the dominant dark matter contribution from QCD axions.

\subsection{Eigenfunctions and eigenvalues}\label{sec:eigen}

\begin{figure} %
\centering
\includegraphics[scale=0.9]{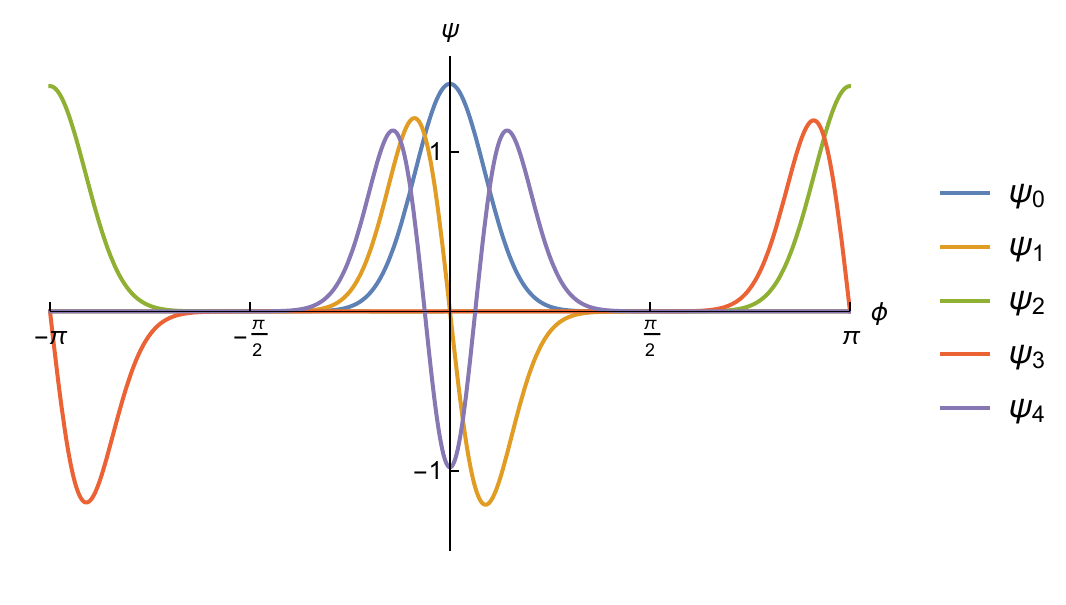}
\caption{The first five eigenfunctions of the eigenvalue equation \eqref{eq: simpleeveq} for $\alpha = 1$, with the ground state $\psi_0$ centred at $\phi = 0$. The parity of the eigenfunction with respect to $\phi = 0$ corresponds to its number.}
\label{fig: alpha1efuncs}
\end{figure}

\noindent 
We found the eigenfunctions and eigenvalues of the full case (\ref{eq: simpleeveq}) numerically using Mathematica.
The first five eigenfunctions for $\alpha = 1$ are represented in Figure \ref{fig: alpha1efuncs}, with $\psi_1$ and $\psi_2$ being degenerate, and likewise for $\psi_3$ and $\psi_4$.  These eigenfunctions enter the spectral coefficients \eqref{eq: coeff-f}, and therefore determine the axion density correlator. 

\begin{figure} %
\centering
\vspace{5pt}
\includegraphics[scale=1.2]{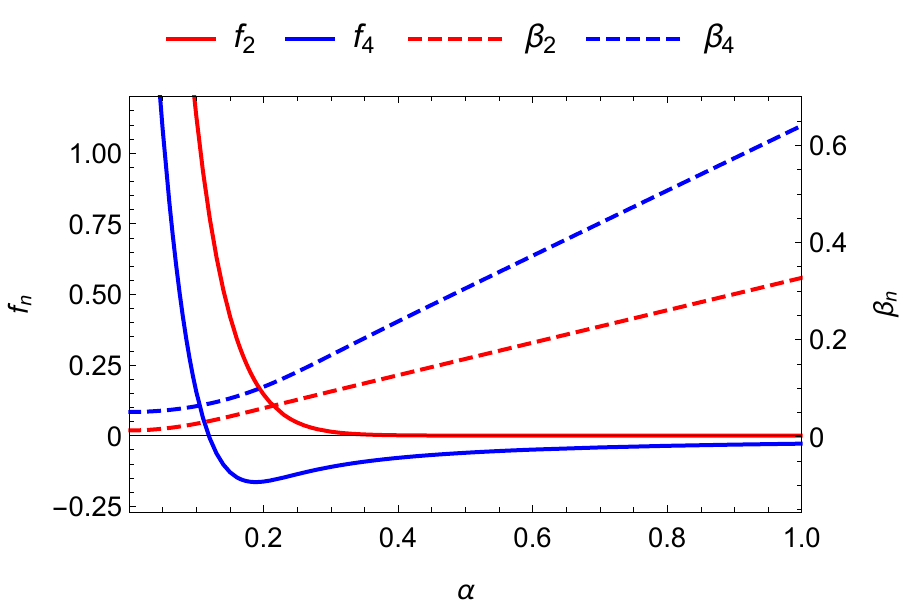}
\caption{The spectral coefficients $f_n$ and the eigenvalues $\beta_n$ contributing to the power spectrum \eqref{equ:PS2} as a function of $\alpha$. The first eigenlevel is shown in red, and the second one is in blue.}
\label{fig: bandfn}
\end{figure}

From Fig. \ref{fig: alpha1efuncs} we observe that the ground state $\psi_0$ is centred at the minimum of the axion potential \eqref{eq: actualaxionpot}, as would be expected since the most likely location of the axion is at rest at the bottom of the potential. However, the higher order eigenfunctions at each degeneracy level are centred on different symmetry points ($\phi = 0$ and $\phi = \pm \pi$), and one is always even and one is odd w.r.t $\phi = 0$. 
The states localised around $\phi=0$, which is the minimum of the potential~(\ref{eq: actualaxionpot}), can be thought of as perturbative states, and are indeed well approximated by the harmonic approximation (\ref{eq: harmapproxevalueeq}) when $\alpha$ is sufficiently large. In contrast, the states localised around $\phi=\pi$ do not have a counterpart in the harmonic spectrum, and correspond physically to non-perturbative states. However, it should be noted that they can still be approximated by a harmonic expansion about $\phi=\pm\pi$.

As our  density function $f(\phi)$ defined in Eq.~\eqref{eq: thefunction} is even, only even eigenfunctions contribute to the spectral expansion (\ref{eq: spatialcorrelator}) of the axion density correlator.
Because the lowest even excited state, $\psi_2$, is non-perturbative, this means that non-perturbative contributions are crucial for the correlation function (\ref{equ:isocorr}), and the standard linear or perturbative approximations are not reliable.

The behaviour of the lowest even eigenvalues $\beta_n$ and the corresponding spectral coefficients $f_n$ as functions of $\alpha$ is shown in Fig.~\ref{fig: bandfn}. 
For large $\alpha$, the eigenvalues increase linearly with $\alpha$, in good agreement with the harmonic approximation (\ref{equ:harmef}). 
Interestingly this is true even for the lowest eigenvalue $\beta_2$ which corresponds to a non-perturbative state and should therefore not be present in the harmonic approximation. In spite of this $\beta_2$ is well approximated by $\beta_1^{\rm harm}$. This agreement of eigenvalues is a consequence of supersymmetry: The supersymmetry transformation (\ref{equ:susy}), combined with a shift of $\phi$, maps $\psi_2$ to $\psi_1$, which is well approximated by $\psi_1^{\rm harm}$.
The next even state, $\psi_4$, is perturbative, and is well approximated by the harmonic eigenfunction $\psi^{\rm harm}_2$ as expected.

For smaller $\alpha$, the eigenvalues $\beta_n$ are no longer well approximated by the harmonic approximation, and instead of vanishing in the limit $\alpha\rightarrow 0$ as the harmonic approximation (\ref{equ:harmef}) predicts, they reach non-zero values. In particular, one finds that
\begin{equation}
    \lim_{\alpha\rightarrow 0}\beta_2=\frac{1}{8\pi^2},
    \label{equ:beta2limit}
\end{equation}
and more generally $\beta_n\ge 1/8\pi^2$ for all even $n$.

From Fig.~\ref{fig: bandfn}, one can also observe that the spectral coefficient $f_2$ becomes very small when $\alpha$ increases. This happens because the spectral coefficients $f_n$ depend on the overlap of the ground state $\psi_0$ with the excited state $\psi_n$. The perturbative states centred around the minimum will  naturally have a higher spectral coefficient (provided that they have the correct parity), while the contribution from non-perturbative states such as $\psi_2$ will depend strongly on the value of $\alpha$ and decreases when $\alpha$ becomes large. Using the analogy with the Schr\"odinger equation, this can be thought of as a tunnelling problem. With higher $\alpha$, the tunnelling between the two minima of $W(\phi)$ becomes exponentially suppressed, and therefore the coefficient $f_2$ becomes exponentially small. 

In contrast, the spectral coefficient $f_4$ of the perturbative state $\psi_4$ behaves as $f_2^{\rm harm}$ at large $\alpha$ and therefore only decreases as $1/\alpha$. On the other hand, it is worth noting that it changes sign at $\alpha\approx 0.12$, and therefore it has a small value in the neighbourhood of that point.

\subsection{Isocurvature power spectrum} \label{isocurvpowerspectrum}

The isocurvature power spectrum (\ref{equ:PS}) at the Planck pivot scale (\ref{equ:Planckpivot}) is
\begin{equation}
\label{equ:PS2}
{\cal P}_S(k_P) \approx \frac{2}{\pi}  \sum_{n=1}^\infty \frac{f_{n}^2}{\langle f \rangle^2} \Gamma\bigg( 2 - 2\frac{H^2}{F_a^2}\beta_{n} \bigg) \sin{\bigg( \frac{H^2}{F_a^2}\pi\beta_{n} \bigg)} e^{-\frac{2H^2N_P}{F_a^2}\beta_{n}},
\end{equation}
where
\begin{equation}
N_P \approx 56 + \frac{1}{2}\ln{\bigg( \frac{H}{8 \times 10^{13} \: \textrm{GeV}} \bigg)}.
\end{equation}

When $H\ll F_a$, the exponent in Eq.~(\ref{equ:PS2}) is small, but the sine factor suppresses the amplitude, and therefore for sufficiently low $H$, the isocurvature constraint is satisfied. This corresponds to the conventional ``anthropic'' stochastic axion region discussed in Refs.~\cite{Hertzberg:2008wr,Wantz:2009it}. 
On the other hand, for $H\gg F_a$ the exponent will always become large and negative because $\beta_2>1/8\pi^2$ as discussed in Section~\ref{sec:eigen},
and therefore the isocurvature amplitude becomes smaller again. This is the ``classic window'' discussed briefly in Ref.~\cite{Hertzberg:2008wr,Wantz:2009it}, and it is the region we focus on.
In this window, the currently observable Universe consists of a large number of regions between which the axion field is uncorrelated, and therefore the expectation value $\langle f\rangle$ in Eq.~(\ref{equ:PS2}) can be identified with the equilibrium expectation value.

\begin{figure} %
\centering
\vspace{23pt}
\includegraphics[scale=1.1]{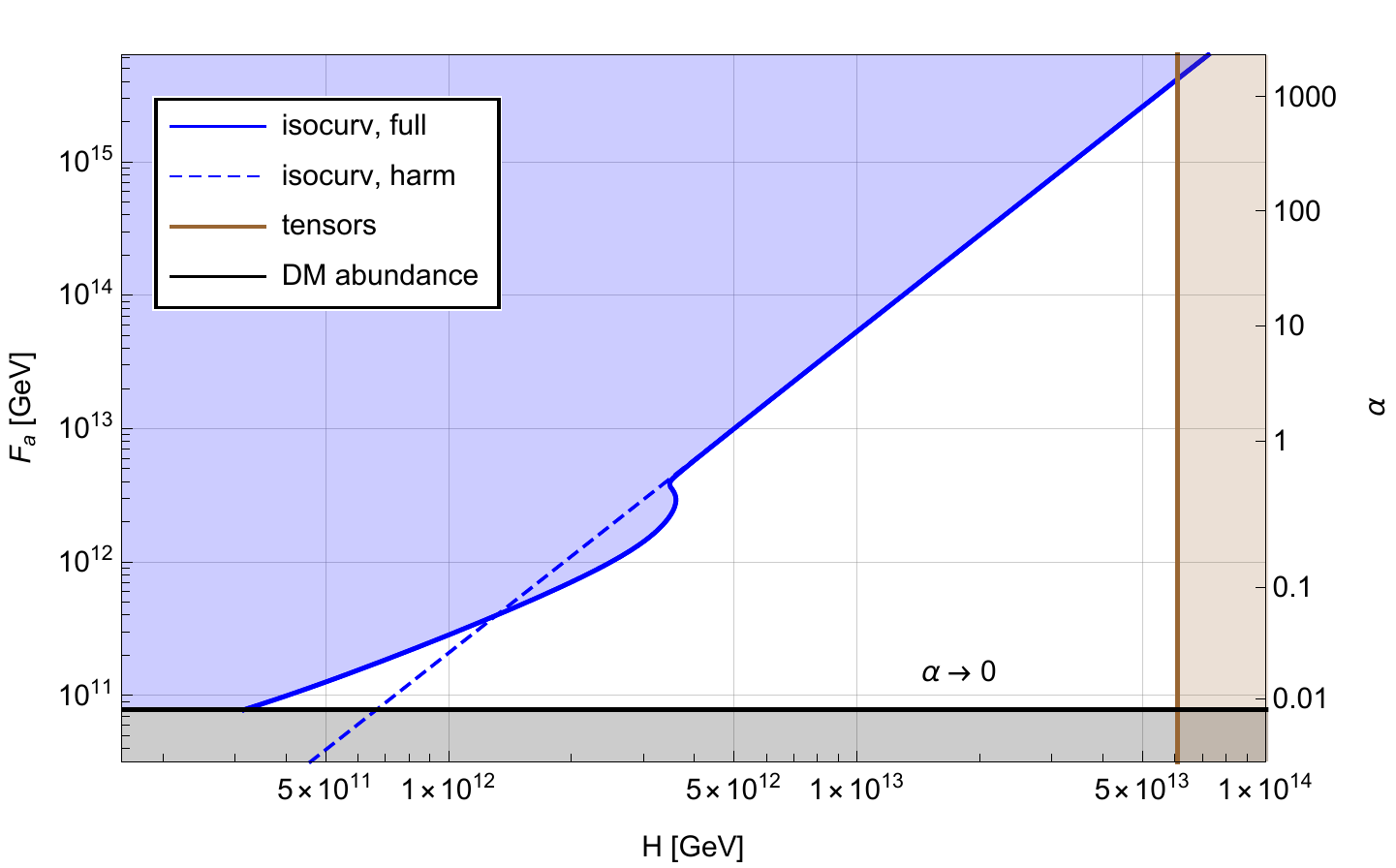}
\caption{Allowed parameter space (white) for the inflationary Hubble parameter $H$ and the axion decay constant $F_a$ in the ``classic window''. 
In the brown region, the tensor mode amplitude would be too high to be compatible with observations. In the black region, the dark matter density would be too low. In the blue region, the isocurvature amplitude is too high. The blue dashed line shows the bound obtained using the harmonic approximation, whereas the solid blue line corresponds to the full stochastic calculation.
The $x$-axis shows the inflationary Hubble rate, 
the left $y$-axis shows the decay constant $F_a$, while the right $y$-axis gives the corresponding value of $\alpha = {V_0}/{H^4}$. 
}
\label{fig: hinfprediction}
\end{figure}

We computed the isocurvature spectrum (\ref{equ:PS2}) as a function of the two free parameters $H$ and $F_a$ using the numerically evaluated eigenvalues $\beta_n$ and eigenfunctions $\psi_n$, and compared the result with the observational bound (\ref{eq: bound}). As expected, the isocurvature spectrum has its maximum values as a function of $H$ between the anthropic and classic windows, and therefore it constrains $H$ from below within the classic window. Other constraints arise from dark matter abundance (\ref{equ:abundance}) and from tensor modes (\ref{equ:Hbound}).
The allowed region is shown in Fig.~\ref{fig: hinfprediction}. The specific case of the QCD axion corresponds to the lower edge of the white region.

Overall, the triangular shape of the allowed region agrees approximately with existing literature~\cite{Hertzberg:2008wr,Wantz:2009it}. However, the lower left corner has a non-trivial shape where the allowed region is slightly narrower for $F_a\gtrsim 5\times 10^{12}~{\rm GeV}$ than the harmonic approximation suggests, and slightly wider for lower values of $F_a$. This is because of the non-linear features captured by the stochastic calculation, specifically the behaviour of the eigenvalues and spectral coefficients shown in Fig.~\ref{fig: bandfn}. 

\begin{figure} %
\centering
\vspace{5pt}
\includegraphics[scale=1.2]{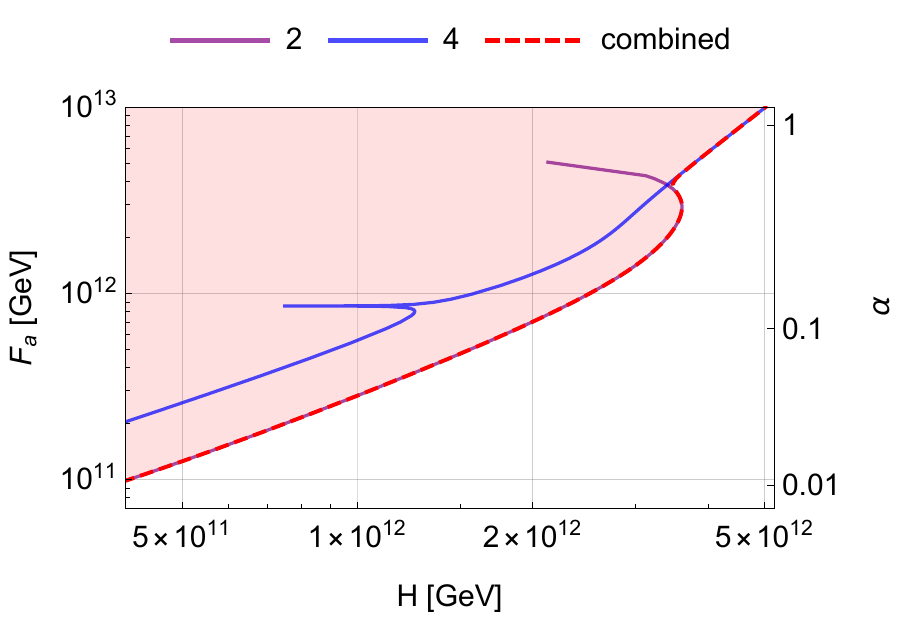}
\caption{
The purple and blue lines show the contribution to the isocurvature bound from the states $n=2$ and $n=4$ in the spectral expansion, respectively. The red line shows the full bound, which includes both contributions.
}\label{fig: transregion}
\end{figure}

For high $F_a$ (corresponding to high $\alpha$), the harmonic approximation works well. Because the spectral coefficient $f_2$ is exponentially small, the dominant term in the sum (\ref{equ:PS2}) corresponds to $\psi_4$, which is a perturbative state well approximated by $\psi_2^{\rm harm}$. However, as illustrated in Fig.~\ref{fig: transregion}, when we move down to $F_a\approx 4 \times 10^{12}~{\rm GeV}$, the spectral coefficient $f_2$ becomes so large that the non-perturbative state $\psi_2$ becomes dominant. This contribution is not included in the conventional linear approximation, which therefore underestimates the isocurvature amplitude. As a result, the allowed region is smaller than the earlier calculations suggested.

Further down, at  $F_a\approx 9\times 10^{11}~{\rm GeV}$, the contribution from $\psi_4$ vanishes temporarily when $f_4$ changes sign. This gives rise to a spike in the blue curve in Fig.~\ref{fig: transregion}, but because the contribution from $\psi_4$ is already subdominant at those values of $F_a$, this feature is not observationally relevant.
At around the same values, the harmonic approximation also fails to describe the eigenvalues $\beta_2$ and $\beta_4$ which instead approach constant values as discussed in Section~\ref{sec:eigen}. This means that the isocurvature spectrum is suppressed relative to the harmonic approximation, and as a consequence the allowed region is larger than what standard linear calculations indicate.

\section{Conclusions} \label{discussion}

In this work, we have used spectral expansions within the stochastic approximation to compute the spectrum of isocurvature perturbations in an axion dark matter model with a cosine potential.
We focused on the ``classic window'' at high Hubble rates $H\gtrsim 10^{12}~{\rm GeV}$ and found, in line with earlier literature, that there is a window of parameters that are compatible with the observed dark matter abundance and the observational constraints on the isocurvature and tensor mode perturbations.

In particular, we found that for axion decay constants in the range $8\times 10^{10}~{\rm GeV}\lesssim F_a \lesssim 5\times 10^{12}~{\rm GeV}$, the isocurvature spectrum is dominated by a non-perturbative contribution and therefore cannot be correctly calculated using standard linear or perturbative techniques. In this range of parameters, the leading term in the spectral expansion does not correspond to small perturbative oscillations of the axion field, but instead to transitions between the minima of the periodic axion potential. In the upper end of the range, this gives an additional contribution to the isocurvature spectrum on top of the perturbative contribution, and therefore the constraints are stronger than the perturbative calculation suggests. At lower values of $F_a$, it is no longer possible to identify the terms in the spectral expansion with perturbative and non-perturbative contributions, and the whole spectrum is more suppressed. Therefore in this region, the constraints are weaker than the perturbative calculation suggests.

Our analysis highlights some interesting features in the behaviour of the theory within the stochastic approximation. Because the theory is invariant under sign change of the potential, the spectrum consists of degenerate pairs of eigenfunctions which have opposite parity and are related to each other by a supersymmetry transformation. When $F_a$ is sufficiently large, one of the eigenfunctions corresponds to a perturbative and the other one to a non-perturbative state. This means that there is a one-to-one mapping between perturbative and non-perturbative contributions and that, in principle, it would be possible to compute non-perturbative effects using perturbative techniques. In our treatment, this follows as a consequence of the stochastic approximation, but it would be interesting to see whether, and to what extent, this correspondence extends to the full quantum field theory.

Finally, we made several simplifying assumptions in our calculations, in particular approximating the inflationary spacetime with the de Sitter spacetime, ignoring any effects of post-inflationary evolution or topological defects such as axion strings, assuming a simple cosine potential for the axion field, as well as using the stochastic approximation itself. These effects are likely to affect the precise numerical values of the bounds, but  at a qualitative level the approximations should all be sound. In particular, the conclusion that non-perturbative effects are crucial for the isocurvature bound on the inflationary Hubble rate in the classic axion window should therefore be valid, but more work is needed to take those effects fully into account.

\acknowledgments{
We would like to thank Eliel Camargo-Molina, Tommi Markkanen and Tommi Tenkanen for useful discussions.
A.R. was supported by the U.K. Science
and Technology Facilities Council grants ST/P000762/1
and ST/T000791/1 and Institute for Particle Physics
Phenomenology Associateship.
}

\bibliographystyle{JHEP}
\bibliography{mybiblio2}

\end{document}